\documentclass[journal ]{new-aiaa}
\usepackage[utf8]{inputenc}
\usepackage{textcomp}

\usepackage{graphicx}
\usepackage{amsmath}
\usepackage[version=4]{mhchem}
\usepackage{siunitx}
\usepackage{longtable,tabularx}
\usepackage{caption}
\usepackage{subcaption}
\usepackage{tikz}
\usepackage{tkz-euclide}

\title{Minimal Differential Lateral Acceleration Configurations for Starshade Stationkeeping in Exoplanet Direct Imaging}

\author{Jackson Kulik\footnote{PhD Student, Center for Applied Math, jpk258@cornell.edu} and Dmitry Savransky \footnote{Assistant Professor, Mechanical and Aerospace Engineering}}

\affil{Cornell University, Ithaca, NY, 14850}
\author{Gabriel J. Soto\footnote{Postdoctoral Researcher, Department of Engineering Physics}}
\affil{University of Wisconsin-Madison, Madison, WI, 53706}

\begin{document}

\maketitle

\section{Introduction}
Exoplanet imaging missions utilizing an external occulter (starshade) for starlight suppression require precise alignment between the telescope and starshade, necessitating maintenance of the starshade orbit during observations. Using differential lateral acceleration between the two spacecraft as a proxy for fuel use and number of required interruptions to the observation, 
we analyze the costs and timing metrics associated with maintaining the starshade in a nominal position, within some tolerance, to block out light from the given star. Our approach is in the same vein as previous analyses from \cite{flinois2020starshade,sirbu2010dynamical,soto2021analytical}. We model the use of impulsive burns to maintain the starshade within a one meter lateral tolerance of the line of sight between telescope and target star \cite{seager2015exo}. Fuel costs and number of stationkeeping maneuvers required during an observation can serve as important considerations in deciding between target stars for observation. These stationkeeping metrics may be used in an objective function that would ideally evaluate quickly as part of a larger scheduling problem optimization such as those described in  \cite{ keithly2020optimal,sanchez2020towards,soto2018optimal,soto2019parameterizing}. 

A stationkeeping strategy that maximizes time between impulsive burns is given by \cite{flinois2020starshade}. This strategy for deadbanding assumes constant differential acceleration between telescope and starshade. The authors of \cite{soto2021analytical} further explored this strategy in a variety of scenarios, analyzing a wider variety of pertinent stationkeeping metrics. Both papers numerically solved ODEs with event driven application of the impulses to find stationkeeping metrics. This is an expensive operation and would not be desirable to carry out as a small part of a larger mission design optimization process, motivating the need for a fast to calculate approximation of the stationkeeping metrics, and a comparison of its accuracy with respect to the methods above. 

We summarize the work of \cite{flinois2020starshade}, and then compare their analytical strategy under ideal conditions against numerical results from \cite{soto2021analytical}. This serves to validate the use of the inexpensive analytical model for stationkeeping. From there, we continue the linear analysis of \cite{flinois2020starshade} on the lateral differential acceleration, focusing on the locations of its minima. We present a geometric picture of differential lateral acceleration, and an analytical expression to approximate these locations associated with low stationkeeping costs. Among starshade positions constrained to the surface of a sphere centered about the telescope, minima of differential lateral acceleration lie on a great circle and its corresponding poles. This intuition and analytical expression for the minima could be used as a heuristic to reduce the design space for scheduling of an exoplanet imaging mission.

\section{Analytical Model of Station Keeping Metrics}
\begin{figure}[ht]
    \centering
    \includegraphics[width=.6\textwidth]{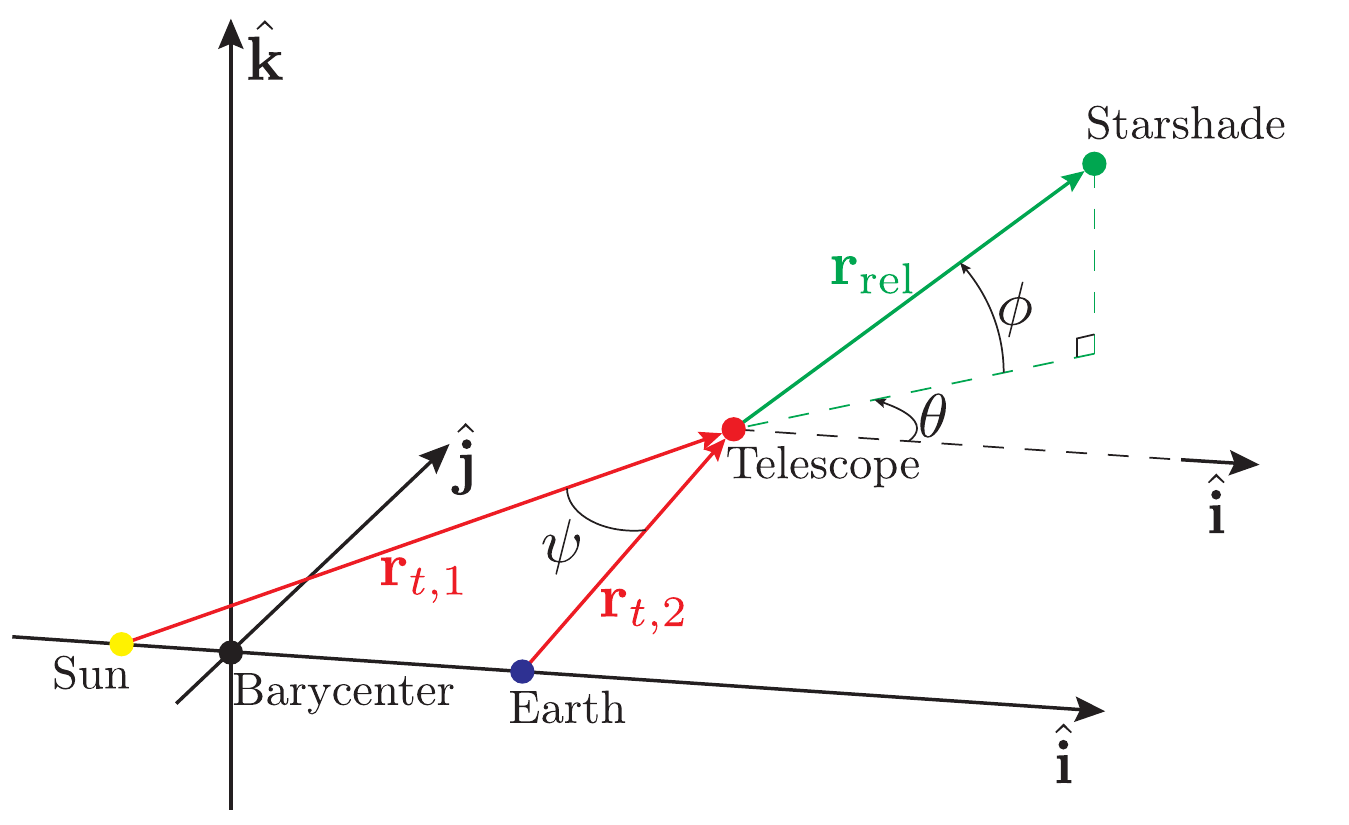}
    \caption{The coordinate frame employed in this paper with relevant bodies and angles depicted.}
    \label{fig:coord_frame}
\end{figure}

We consider the inertial accelerations of the telescope $\mathbf{a}_t$ and starshade $\mathbf{a}_s$ with respect to the inertially non-accelerating barycenter at an initial epoch, and then assume that the differences in inertial accelerations between satellites remain constant over the course of the observation. In the inertial frame, this maintains the direction of the relative position vector $\mathbf{r}_\mathrm{rel}$ between the starshade at $\mathbf{r}_s$ and telescope at $\mathbf{r}_t$, preserving alignment of the telescope, starshade, and target. For convenience, consider an inertial frame which happens to be aligned with the CR3BP rotating frame at the initial epoch and in units of AU. The frame is depicted in Fig. \ref{fig:coord_frame} with the Earth-Sun barycenter located at the origin. Consider all accelerations from this point forward as inertial.

In the inertial reference frame, the acceleration of satellite $j=t,s$ from two bodies is given by

\begin{align}
    \mathbf{a}_j&=\mathbf{a}_{j,1}+\mathbf{a}_{j,2}\\
    \mathbf{a}_{j,1}&=\frac{(\mu-1)(\mathbf{r}_j+\mu\hat{\mathbf{i}})}{||\mathbf{r}_j+\mu\hat{\mathbf{i}}||^3}\\
    \mathbf{a}_{j,2}&=\frac{-\mu\big(\mathbf{r}_j+(\mu-1)\hat{\mathbf{i}}\big)}{||\mathbf{r}_j+(\mu-1)\hat{\mathbf{i}}||^3}
\end{align}
where $\mu$ is the mass parameter, subscript 1 refers to the primary body (the Sun), 2 refers to the secondary body (the Earth), $\hat{\mathbf{i}}$ is the unit vector in the Sun-Earth direction, and $\mathbf{r}$ refers to position. The Sun is assumed to be at $-\mu\hat{\mathbf{i}}$ and the Earth at $(1-\mu)\hat{\mathbf{i}}$.

The differential acceleration between starshade and telescope is given by the difference in the inertial second time derivatives of position 
\begin{equation}
    \delta\mathbf{a}=\mathbf{a}_s-\mathbf{a}_t=\ddot{\mathbf{r}}_s-\ddot{\mathbf{r}}_t.
\end{equation}
Starshade position is nominally given by
\begin{equation}
    \mathbf{r}_s=\mathbf{r}_t+R(\cos\phi\cos\theta\hat{\mathbf{i}}+\cos\phi\sin\theta\hat{\mathbf{j}}+\sin\phi\hat{\mathbf{k}})
\end{equation}
Where, as seen in Fig. \ref{fig:coord_frame}, $R=\Vert \mathbf{r}_\mathrm{rel}\Vert$ is the nominal telescope-starshade separation distance, $\theta$, the ecliptic longitude, is the in-plane angle of the star from the Sun-Earth vector at the starting epoch, and $\phi$, the ecliptic latitude, is the out-of-plane angle measured from the ecliptic plane as measured with respect to the starshade location.

Consider the axial and lateral components of the differential acceleration, where axial denotes the direction of the telescope line of sight to the target star, and lateral denotes the component orthogonal to the line of sight. The lateral differential acceleration magnitude is given by:

\begin{equation}
    \delta a_l=||\delta\mathbf{a}-(\delta\mathbf{a}\cdot\hat{\mathbf{r}}_\mathrm{rel})\hat{\mathbf{r}}_\mathrm{rel}|| \,,
\end{equation}
where $$\hat{\mathbf{r}}_\mathrm{rel}=\cos\phi\cos\theta\hat{\mathbf{i}}+\cos\phi\sin\theta\hat{\mathbf{j}}+\sin\phi\hat{\mathbf{k}}$$

The starshade has a nominal relative position with respect to the telescope and a lateral tolerance $r_\mathrm{tol}$ from this position. We consider lateral stationkeeping, disregarding axial stationkeeping, because axial tolerances are typically larger than the distances traversed during a typical observation \cite{soto2021analytical}.

In the optimal strategy outlined in \cite{flinois2020starshade}, the starshade begins some nominal axial distance from the telescope and at the edge of the $r_\mathrm{tol}$ lateral tolerance disc in the direction of lateral relative acceleration. The starshade is initialized with lateral relative velocity chosen in the opposite direction of lateral relative acceleration. The magnitude of the velocity is chosen such that the starshade will reach the exact opposite end of the disk with radius given by the lateral tolerance, and then come back down to its initial position under the constant lateral acceleration. At this point, the starshade applies an impulsive burn to repeat the process. The resulting stationkeeping metrics are given as follows:
\begin{align}
\label{metric1}
    T&=4\sqrt{\frac{r_\mathrm{tol}}{\delta a_l}}\\
    N&=\text{floor}\bigg(\frac{\tau_\mathrm{obs}\sqrt{\delta a_l}}{4\sqrt{r_\mathrm{tol}}}\bigg)\\
    \Delta \text{v}&=4N\sqrt{\delta a_l r_\mathrm{tol}} \,,
    \label{metric2}
\end{align}
where $\tau_\mathrm{obs}$ is the overall observation/stationkeeping time, $T$ is the amount of time between impulsive burns, $N$ is the number of burns after the initial epoch (not counting the initialization of the starshade), and $\Delta\mathrm{v}$ is the cost in terms of impulsive velocity changes to perform the stationkeeping. These results were presented in nondimensional form in more generality for arbitrary initialization positions on the tolerance disk \cite{flinois2020starshade}. Notice that $\Delta\mathrm{v}$ is largely independent of the stationkeeping tolerance, and varies as $\tau \delta a_l$, the product of observation time and acceleration magnitude. We can obtain various other metrics as a function of these as in \cite{soto2021analytical}.



\begin{figure}[ht]
     \centering
\begin{picture}(400,180)
\centering
\put(0,0){ \includegraphics[height=180pt]{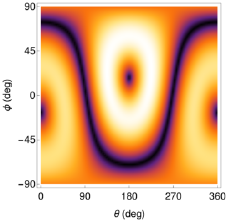}}
\put(170,0){ \includegraphics[height=180pt]{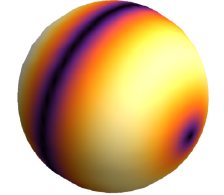}}
\put(370,20){ \includegraphics[height=120pt]{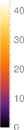}}
\put(350,190){\begin{minipage}[t]{50pt}
\begin{align*}
\delta a_l (\mu \mathrm{m}/\mathrm{s}^2)
\end{align*}
 \end{minipage}}
\end{picture}

\caption{The differential lateral acceleration in $\mu\mathrm{m}/\mathrm{s}^2$ as a function of target coordinates. The density plot on the left is accompanied by its projection onto the unit sphere on the right.}

\label{fig:acc_plots_sphere}
\end{figure}

\begin{figure}[ht]
\centering
     \begin{subfigure}[b]{0.33\textwidth}
         \centering
         \includegraphics[width=\textwidth]{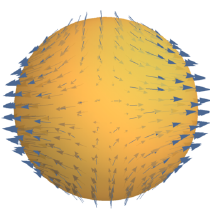}
     \end{subfigure}
          \begin{subfigure}[b]{0.33\textwidth}
         \centering
         \includegraphics[width=\textwidth]{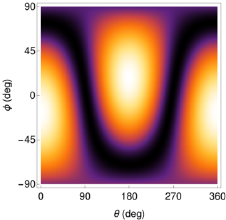}
     \end{subfigure}
               \begin{subfigure}[b]{0.05\textwidth}
         \centering
         \includegraphics[width=\textwidth]{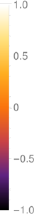}
     \end{subfigure}
    \caption{On the left, the differential acceleration vector field along the unit sphere about the telescope. 
    On the right, the cosine of the angle between relative position and differential acceleration.}
    \label{fig:vec_field}
\end{figure}

These metrics are relatively simple to evaluate compared to the numerical solution of an ODE that depends on repeatedly evaluating the differential acceleration. As each of these metrics depends simply on differential lateral acceleration, we study this quantity as a proxy. A plot of lateral differential acceleration is given in Fig. 8 of \cite{flinois2020starshade}. However, only a small region around a maximum is depicted as worst case scenarios were the primary focus of the work. The overall structure of the differential acceleration is not given, and in particular, the minima are not shown. Fig. \ref{fig:acc_plots_sphere} sheds some light on the structure of differential lateral acceleration and resembles the color plots seen in Fig. 9 of \cite{soto2021analytical}. Looking at the same data plotted over the unit sphere centered about the telescope, we can understand that the minima of lateral acceleration magnitude lie on a great circle and its corresponding poles. In Fig. \ref{fig:acc_plots_sphere}, the telescope is assumed to be at $(x,y,z)=(1+2.5/150,0,1/150)$. The starshade is assumed to be $R=100,000$ kilometers away from the telescope. The existence of the great circle and poles can be explained by the rejection of the axial component of acceleration: Fig. \ref{fig:vec_field} shows a plot of the differential acceleration vector field, demonstrating that the vector field is orthogonal to the sphere in the axial or anti-axial direction along the same great circle and at the two corresponding poles. More specifically, the vector field in Fig. \ref{fig:vec_field} points outwards on the far left and right, and then inwards towards the origin along the circle lying between the two extreme points. We also depict the cosine of the angle between relative position and differential acceleration in Fig. \ref{fig:vec_field}. This gives a less intuitive but more clear representation of the direction of the differential acceleration vector. Values of $-1,0,1$ correspond to differential acceleration that is purely anti-axial, lateral, and axial respectively.  

\begin{figure}[ht]
     \centering
\begin{picture}(400,180)
\centering
\put(-40,0){ \includegraphics[height=120pt]{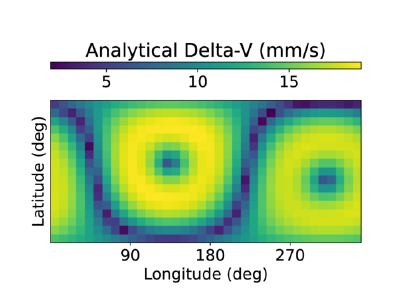}}
\put(110,0){ \includegraphics[height=120pt]{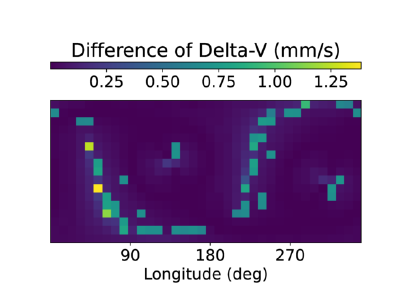}}
\put(260,0){ \includegraphics[height=120pt]{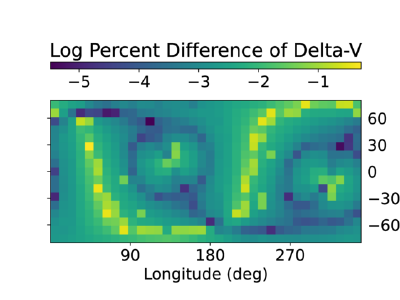}}
\end{picture}

\caption{Comparison between numerical and analytical models of the average delta-v to perform a single stationkeeping impulse in mm/s during an observation at 340 days from the start of the reference halo orbit.}
        \label{fig:dif_anal_numer}
\end{figure}

Figure \ref{fig:dif_anal_numer} demonstrates the similarity between the numerically computed and analytically computed $\Delta \mathrm{v}$ metric for a telescope $340$ days from initialization of the halo orbit. Analytical computation is described above, while the numerical computation is carried out with only gravitational effects as in \cite{soto2021analytical}. We see that the two metrics agree very well except along the minima of $\Delta \mathrm{v}$, where they differ by up to half of the analytical value. The effects of nonconstant differential lateral acceleration are a plausible explanation of these differences. It is reasonable to believe that at these minima induced by the direction of differential acceleration, small changes in differential acceleration direction would have greater effects than at other locations.  However, the overall structure from the two approaches is the same, with minima in the nearly the same locations.


\section{Approximate Minima of Lateral Acceleration}

\begin{figure}[ht]
    \centering
    \includegraphics[width=.6\textwidth]{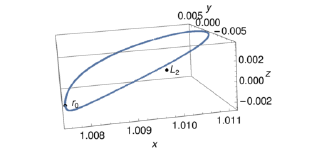}
    \caption{The 6 month period reference halo orbit, with the initial position labeled in canonical units of the CR3BP.}
    \label{fig:halo}
\end{figure}

\begin{figure}[ht]
     \centering
     \begin{subfigure}[b]{0.33\textwidth}
         \centering
         \includegraphics[width=\textwidth]{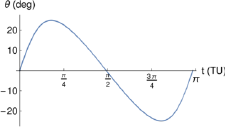}
     \end{subfigure}
     \begin{subfigure}[b]{0.33\textwidth}
         \centering
         \includegraphics[width=\textwidth]{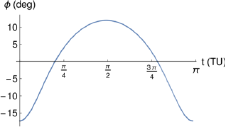}
     \end{subfigure}\\
    \begin{subfigure}[b]{0.32\textwidth}
         \centering
         \includegraphics[width=\textwidth]{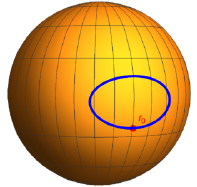}
     \end{subfigure}
        \caption{Pole location as the telescope moves along the reference halo orbit trajectory in terms of ecliptic longitude, latitude and location on the unit sphere.
        }
        \label{fig:poles_over_orb}
\end{figure}

The location of one of the poles shown in Fig. \ref{fig:acc_plots_sphere} can be computed by numerical minimization. Along our six month reference halo orbit depicted in Fig. \ref{fig:halo}, we give the angular locations of the pole in Fig. \ref{fig:poles_over_orb}.

Employing a poor initial guess can cause numerical methods to identify minima along the great circle rather than the pole. Serving as an initial guess for numerical minimization, a quick approximation of the pole's location can be obtained from the eigenvectors of the Jacobian matrix $D$ of the inertial acceleration vector of the telescope.

\begin{equation}
    \delta\mathbf{a}(\mathbf{r}_t)= \mathbf{a}(\mathbf{r}_t+\mathbf{r}_\mathrm{rel})-\mathbf{a}(\mathbf{r}_t) \approx D \mathbf{a}(\mathbf{r}_t)\cdot\mathbf{r}_\mathrm{rel} \label{eqn:eigen}
\end{equation}

Equation \ref{eqn:eigen} demonstrates that differential acceleration will be completely axial in the corresponding linear system precisely when the differential acceleration $\delta\mathbf{a}(\mathbf{r}_t)$ is parallel or antiparallel to the offset vector between telescope and starshade $\mathbf{r}_\mathrm{rel}$. This is the case when $\mathbf{r}_\mathrm{rel}$ is an eigenvector of $D \mathbf{a}(\mathbf{r}_t)$. Given the structure of the vector field in Fig. \ref{fig:vec_field}, the eigenvector corresponding to the pole will be the one corresponding to a positive eigenvalue. The other two eigenvectors correspond to points on the great circle. In the case of our reference halo orbit, the difference in angular position of the pole between the numerical method in Fig. \ref{fig:poles_over_orb} and the linear approximation is on the order of a degree for $R=100,000$km, scaling linearly with $R$.

Even with the relatively easy to compute and robust eigenvector interpretation, we still seek an analytical expression for the pole locations. We begin by approximating the differential acceleration by a binomial series as in \cite[][page 391]{vallado2001fundamentals}.

\begin{align}
\label{eqn:approx_da}
\delta\mathbf{a}&=\delta\mathbf{a}_1+\delta\mathbf{a}_2\\
\delta\mathbf{a}_j&\approx \frac{a_{t,j}\mathbf{r}_\mathrm{rel}}{r_{t,j}}-\frac{3\mathbf{a}_{t,j}}{r_{t,j}}(\hat{\mathbf{a}}_{t,j}\cdot\mathbf{r}_\mathrm{rel}) \label{eqn:approx_diff_acc}\\
\delta a_{j,l}&\approx \frac{3a_{t,j}\sin\theta_j}{r_{t,j}}(\hat{\mathbf{a}}_{t,j}\cdot\mathbf{r}_\mathrm{rel})=\frac{3Ra_{t,j}\sin\theta_j\cos\theta_j}{r_{t,j}} \,,
\end{align}
where $j$ denotes the relevant body (1 for Sun and 2 for Earth), the subscript $l$ denotes the lateral component of the acceleration and $\theta_j$ the angle between $\mathbf{r}_\mathrm{rel}$ and $\mathbf{a}_{t,j}$ or equivalently between $\mathbf{r}_\mathrm{rel}$ and $-\mathbf{r}_{t,j}$. We will use $a,r$ to denote the norms $\left\Vert \mathbf{a}\right\Vert,\left\Vert \mathbf{r}\right\Vert$ respectively. In order to find configurations that induce axial differential acceleration, we find values of $\mathbf{r}_\mathrm{rel}$ that balance $\delta\mathbf{a}_1$ against $\delta\mathbf{a}_2$ so that their sum is in the direction of $\mathbf{r}_\mathrm{rel}$. This corresponds to when their lateral components $\delta a_{1l}$ and $\delta a_{2l}$ are equal as shown in Fig. \ref{fig:drawing1}. Since the first term in the approximate differential acceleration from each body is in the direction of $\mathbf{r}_\mathrm{rel}$, the problem reduces to balancing the second terms of Equation \ref{eqn:approx_diff_acc}. After dividing by common terms, axial differential acceleration is equivalent to the following conditions:

\begin{figure}[ht]
\centering
\scalebox{.85}{
\begin{tikzpicture}
\draw[thick,->] (0,0) -- (4,0) node[anchor=north east] {$\mathbf{r}_\mathrm{rel}$};
\tkzDefPoint(0,0){T}  \tkzLabelPoint[left](T){Telescope}

\tkzDrawPoints[fill=black](T)

\draw[thick,->] (0,0) -- (6,-4) node[anchor=north east] {$\mathbf{\delta a}_1$};
\draw[thick,->] (0,0) -- (10,4) node[anchor=south east] {$\mathbf{\delta a}_2$};




\draw[dashed] (4,0) -- (10,0);
\draw[thick] (10,0) -- (10,4);
\draw[thick] (6,0) -- (6,-4);
\draw (6,-2) node[rotate=90] {$||$};
\draw (5.5,-2.5) node[] {$\delta a_{1l}$};
\draw (10,2) node[rotate=90] {$||$};
\draw (9.5,1.5) node[] {$\delta a_{2l}$};

\end{tikzpicture}
}

\caption{The desired orientation of the telescope to starshade vector ($\mathbf{r}_\mathrm{rel}$) relative to differential acceleration vectors from Sun and Earth gravity ($\delta\mathbf{a}_1,\delta\mathbf{a}_2$). 
} \label{fig:drawing1}
\end{figure}
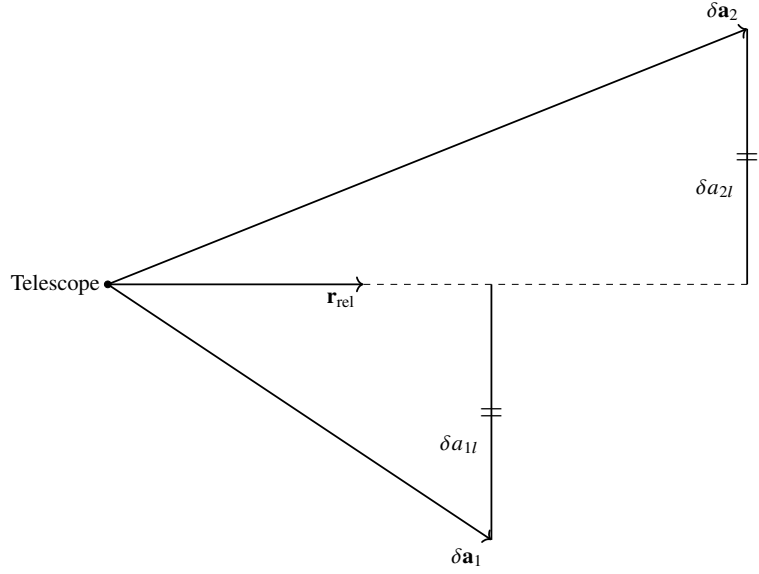

\begin{align}
        \delta a_{1,l}&=\delta a_{2,l} &\iff\\
        0&=\frac{a_{t,1}}{r_{t,1}}\cos\theta_1\sin\theta_1-\frac{a_{t,2}}{r_{t,2}}\cos\theta_2\sin\theta_2 &\iff\\
        0&=\frac{a_{t,1}}{r_{t,1}}\sin2\theta_1-\frac{a_{t,2}}{r_{t,2}}\sin2\theta_2 \,. \label{eqn:trigtermprev}
\end{align}

Defining $\psi=\theta_1+\theta_2$ as the angle between the vectors from the telescope to the two bodies, such that:
\begin{equation}
\psi=\arccos\bigg(\frac{r_{t,1}^2+r_{t,2}^2-1}{2r_{t,1}r_{t,2}}\bigg) \,,
\end{equation}
equation \ref{eqn:trigtermprev} becomes:
\begin{equation}
0=\frac{a_{t,1}}{r_{t,1}}\sin2\theta_1-\frac{a_{t,2}}{r_{t,2}}\sin(2\psi-2\theta_1) \label{eqn:trigterm}
\end{equation}
Expanding the trigonometric term in equation \ref{eqn:trigterm}, we obtain
\begin{align}
\theta_1&=\frac{1}{2}\arctan\bigg(\frac{\sin(2\psi)a_2/r_2}{a_1/r_1+\cos(2\psi)a_2/r_2}\bigg)
\label{eqn:theta}
\end{align}
dropping the subscript $t$ for readability.

To summarize, the location of one of the poles is in the plane of the Earth, Sun, and telescope, situated between the telescope to Earth vector ($\mathbf{r_{t,2}}$) and the telescope to Sun vector ($\mathbf{r}_{t,1}$), $\theta_1$ in angle away from the telescope to Sun vector, $\mathbf{r}_{t,1}$. One may rotate a unit vector in the direction from the telescope to the Sun, $\hat{\mathbf{r}}_{t,1}$, about the $-z_t\hat{\mathbf{j}}+y_t\hat{\mathbf{k}}$ axis by $\theta_1$ to obtain a unit vector in the direction of the pole. Note that the results of this approach are identical to the eigenvector approach as well as numerical minimization of the approximate linear differential lateral acceleration. Given the location of a single pole that we have described, the other pole and the corresponding great circle are also known.

\begin{figure}[ht]
     \centering
\begin{picture}(400,180)
\centering
\put(0,0){ \includegraphics[height=180pt]{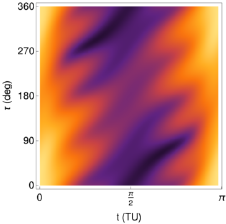}}
\put(190,0){ \includegraphics[height=178pt]{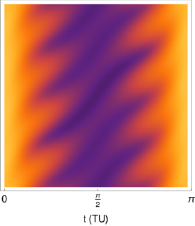}}
\put(360,30){ \includegraphics[height=120pt]{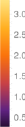}}
\put(350,190){\begin{minipage}[t]{50pt}
\begin{align*}
\delta a_l (\mu \mathrm{m}/\mathrm{s}^2)
\end{align*}
 \end{minipage}}
\end{picture}

        \caption{Differential lateral acceleration in $\mu\mathrm{m}/\mathrm{s}^2$ is depicted along the great circle for a telescope moving along the reference halo orbit. 
        Exact (left) and approximate (right) pole positions describe the great circle.}
        \label{fig:acc_plots_gc}
\end{figure}

The great circle is of particular interest since it is a higher dimensional object than the two poles. Many more stars will be easily observable along the great circle than on or near the poles. As a result, it becomes important to quantify how low the differential lateral acceleration is along the great circle. We employ equation \ref{eqn:theta} to find a unit vector in the approximate direction of the pole and potentially improve on this by numerical minimization. Upon obtaining the approximate or the exact pole location, the corresponding great circle consists of all of the unit vectors orthogonal to the pole vector. We may parameterize this great circle arbitrarily with a parameter $\tau$.  Figure \ref{fig:acc_plots_gc} presents the differential lateral acceleration along the great circles corresponding to the exact and approximate pole locations at each point along the halo orbit. Position of the telescope along the halo orbit is given along the x-axis by the time $t$ from the initial epoch when the telescope is located at $\mathbf{r}_0$ as shown in Fig. \ref{fig:halo}. At each of these values of $t$, the position of the telescope leads to a direction for a pole given by equation \ref{eqn:theta}, which correspond to the values in Fig. \ref{fig:poles_over_orb}. Each time $t$ maps to a pole location, which in turn maps to a single corresponding great circle.  Position along the great circle is given on the y-axis by $\tau$, the arbitrarily chosen phasing parameter. Note that time values near $0$ and $\pi$ lead to generally higher differential lateral accelerations as the two satellites are closer to the Earth at this point in the halo orbit. One can see that as the telescope moves along its halo orbit, even at its closest point to Earth, the differential lateral acceleration along the corresponding great circle is consistently one to two orders of magnitude below the highest values shown in Fig. \ref{fig:acc_plots_sphere}.  Using equations \ref{metric1}-\ref{metric2}, with a lateral position tolerance of 1 meter, this translates to a difference between six stationkeeping maneuvers per hour in the worst case and fewer than one interruption per hour for observations along the great circle. Considering both the exact or approximate great circle effectively determines convenient choices of stars for more fuel efficient and less frequently interrupted observations.


\section{Conclusion}
We have explored trends involving station keeping metrics for observation of exoplanets with a telescope and starshade pair. Differential lateral acceleration proved to be an effective proxy for metrics such as time between interruptions for starshade orbit maintenance, and delta-v. For a fixed position of the telescope, differential lateral acceleration as a function of the line-of-sight vector from telescope to target star proved to possess minima along two poles and their corresponding great circle. We derived an analytical approximation for the location of these minima, and demonstrated that for a design reference mission, observing stars along the great circle yields magnitudes of differential lateral acceleration that are one or two orders of magnitude lower than if the target were chosen indiscriminately. We advocate cognizance of this great circle of easy to observe targets during design of exoplanet missions. By considering targets as they cross this great circle, one may potentially pare down the large design space for scheduling problems related to exoplanet imaging missions.

\section*{Funding Sources}
This work was supported by NASA JPL SURP grant RSA No. 1644208. 

\bibliography{main}

\begin{thebibliography}{9}
\newcommand{\enquote}[1]{``#1''}
\providecommand{\natexlab}[1]{#1}
\providecommand{\url}[1]{\texttt{#1}}
\providecommand{\urlprefix}{URL }
\expandafter\ifx\csname urlstyle\endcsname\relax
  \providecommand{\doi}[1]{\discretionary{}{}{}https://doi.org/#1}\else
  \providecommand{\doi}[1]{\discretionary{}{}{}\urlstyle{rm}\url{https://doi.org/#1}}\fi

\bibitem[{Flinois et~al.(2020)Flinois, Scharf, Seubert, Bottom, and
  Martin}]{flinois2020starshade}
Flinois, T.~L., Scharf, D.~P., Seubert, C.~R., Bottom, M., and Martin, S.~R.,
  \enquote{Starshade formation flying II: formation control,} \emph{Journal of
  Astronomical Telescopes, Instruments, and Systems}, Vol.~6, No.~2, 2020, p.
  029001.

\bibitem[{Sirbu et~al.(2010)Sirbu, Karsten, and Kasdin}]{sirbu2010dynamical}
Sirbu, D., Karsten, C.~V., and Kasdin, N.~J., \enquote{Dynamical performance
  for science-mode stationkeeping with an external occulter,} \emph{Space
  Telescopes and Instrumentation 2010: Optical, Infrared, and Millimeter Wave},
  Vol. 7731, International Society for Optics and Photonics, 2010, p. 773152.

\bibitem[{Soto et~al.(2021)Soto, Savransky, and Morgan}]{soto2021analytical}
Soto, G.~J., Savransky, D., and Morgan, R., \enquote{Analytical model for
  starshade formation flying with applications to exoplanet direct imaging
  observation scheduling,} \emph{Journal of Astronomical Telescopes,
  Instruments, and Systems}, Vol.~7, No.~2, 2021, p. 021209.

\bibitem[{Seager et~al.(2015)Seager, Turnbull, Sparks, Thomson, Shaklan,
  Roberge, Kuchner, Kasdin, Domagal-Goldman, Cash et~al.}]{seager2015exo}
Seager, S., Turnbull, M., Sparks, W., Thomson, M., Shaklan, S.~B., Roberge, A.,
  Kuchner, M., Kasdin, N.~J., Domagal-Goldman, S., Cash, W., et~al.,
  \enquote{The Exo-S probe class starshade mission,} \emph{Techniques and
  Instrumentation for Detection of Exoplanets VII}, Vol. 9605, International
  Society for Optics and Photonics, 2015, p. 96050W.

\bibitem[{Keithly et~al.(2020)Keithly, Savransky, Garrett, Delacroix, and
  Soto}]{keithly2020optimal}
Keithly, D.~R., Savransky, D., Garrett, D., Delacroix, C., and Soto, G.,
  \enquote{Optimal scheduling of exoplanet direct imaging single-visit
  observations of a blind search survey,} \emph{Journal of Astronomical
  Telescopes, Instruments, and Systems}, Vol.~6, No.~2, 2020, p. 027001.

\bibitem[{Sanchez(2020)}]{sanchez2020towards}
Sanchez, W.~D., \enquote{Towards fuel-efficient formation flying of an
  observatory and external occulter at Sun-Earth L2,} Ph.D. thesis,
  Massachusetts Institute of Technology, 2020.

\bibitem[{Soto et~al.(2018)Soto, Keithly, Garrett, Delacroix, and
  Savransky}]{soto2018optimal}
Soto, G., Keithly, D., Garrett, D., Delacroix, C., and Savransky, D.,
  \enquote{Optimal starshade observation scheduling,} \emph{Space Telescopes
  and Instrumentation 2018: Optical, Infrared, and Millimeter Wave}, Vol.
  10698, International Society for Optics and Photonics, 2018, p. 106984M.

\bibitem[{Soto et~al.(2019)Soto, Savransky, Garrett, and
  Delacroix}]{soto2019parameterizing}
Soto, G.~J., Savransky, D., Garrett, D., and Delacroix, C.,
  \enquote{Parameterizing the search space of starshade fuel costs for optimal
  observation schedules,} \emph{Journal of Guidance, Control, and Dynamics},
  Vol.~42, No.~12, 2019, pp. 2671--2676.

\bibitem[{Vallado(2001)}]{vallado2001fundamentals}
Vallado, D.~A., \emph{Fundamentals of astrodynamics and applications}, Vol.~4,
  Springer Science \& Business Media, 2001.

\end{thebibliography}


\end{document}